\documentclass[sort&compress
    ,final            
 ]
  {aipproc}

\layoutstyle{6x9}

\newcommand{\be}{\begin{equation}}
\newcommand{\ee}{\end{equation}}
\newcommand{\bea}{\begin{eqnarray}}
\newcommand{\eea}{\end{eqnarray}}
\newcommand{\open}{{<\kern -0.3 em{\scriptscriptstyle )}}}

\begin{document}

\title{Transversity}


\author{Umberto D'Alesio}{
  address={Dipartimento di Fisica, Universit\`a di Cagliari, Cittadella Universitaria, and
  Istituto Nazionale di Fisica Nucleare, Sezione di Cagliari, C.~P.~170, I-09042 Monserrato (CA), Italy}
}

\begin{abstract}
Transversity distribution, together with the unpolarized and the helicity distributions, represents a basic piece of information on the internal structure of nucleons. Its peculiar property of being a chiral-odd quantity implies that it can be accessed only in combination with another chiral-odd partner, making it much harder to measure. In this talk I will review its properties and describe the most relevant and feasible ways to access it. Recent phenomenological extractions, their highlights and limitations, as well as perspectives are discussed.
\end{abstract}

\maketitle


\section{General properties}

In a collinear framework (or if the intrinsic transverse momenta are integrated over) three building blocks characterize completely the internal structure of a fast moving nucleon: the momentum distribution ($q(x)$), the helicity distribution ($\Delta q(x)$) and the transversity distribution ($\Delta _Tq(x)$ or $h_1^q(x)$)~\cite{Barone:2001sp}.
On equal footing on the theory side, the first two have been analyzed experimentally for decades and considerable information, even if with different accuracy, has been gathered on them. By contrast, $\Delta_Tq$, while extensively studied in a large class of models, only recently has been accessed phenomenologically.

In a partonic picture its probabilistic interpretation is straightforward: given a transversely polarized nucleon (w.r.t.~its direction of motion) and denoting with $q_{\uparrow\downarrow}$ the number density of quarks with polarization parallel (antiparallel) to that of the nucleon, the transversity is simply the difference $q_\uparrow-q_\downarrow$. More formally, in a field theoretical language, $\Delta_Tq(x)$ is given in terms of a hadronic matrix element of a nonlocal operator ($\sim \langle \bar\psi(0)i\sigma^{1+}\gamma_5 \psi(\xi^-) \rangle$) and is a leading-twist quantity, like $q(x)$ and $\Delta q(x)$.

Another way of looking at this function is in terms of the forward quark-nucleon amplitude, $F_{\Lambda\Lambda'}^{\lambda\lambda'}$, where $\Lambda$ ($\lambda$) are the nucleon (quark) helicities:  transversity is then given as $F_{+-}^{+-}$, showing that in the helicity basis it is not related to (a difference of) probabilities, but to the interference of amplitudes. Off-diagonal in the quark-nucleon helicities, it is a chiral-odd function,  to be compared with the chiral-even unpolarized and longitudinally polarized distributions, given respectively by the sum and the difference of $F_{++}^{++}$ and $F_{++}^{--}$.
This explains why transversity escaped, experimentally, for a long time. Being a chiral-odd quantity, in fact, it cannot be accessed via the helicity (chirality) conserving inclusive deep inelastic scattering (DIS) processes. In order to measure $\Delta_Tq$, the chirality must be flipped twice, hence another chiral-odd partner is required.

Transversity does not couple to gluons, implying its non-singlet $Q^2$-evolution (and its strong suppression at low $x$) and no gluon transversity exists for a spin-1/2 hadron.
For relativistic quarks $\Delta_T q$ is expected to be different from $\Delta q$, opening a window on the relativistic motion of quarks in a nucleon. It obeys a nontrivial bound~\cite{Soffer:1994ww}, namely $|\Delta_T q|\le (q +\Delta q$)/2, and it is the only source of information on the tensor charge ($\int (\Delta_T q - \Delta_T \bar q)$), a fundamental charge (calculable on the lattice~\cite{Gockeler:2006zu}), as important as the vector ($\int (q - \bar q)$) and the axial ($\int (\Delta q + \Delta \bar q)$) charges.

\section{Phenomenology}

The requirement of a chiral-odd partner implies that one has to look at processes involving at least two hadrons.
We can consider two classes of observables: 1) the simplest case, from the theory side, is a double transverse spin asymmetry (DtSA), where the chiral-odd partner is a second transversely polarized hadron, either in the initial (a) or in the final (b) state; 2) another (and fruitful) case is a single spin asymmetry (SSA), either within the approach based on transverse momentum dependent distributions (TMDs), encoding correlations between spin and intrinsic transverse momenta ($k_\perp$), either in a framework based on dihadron fragmentation functions (DiFFs), encoding the interference between different partial waves of a dihadron system.
Other cases, not discussed here, are possible, like higher-twist functions or higher-spin particles (like $\rho$ mesons).

Among all the above options only one (1a) requires a second polarized beam, i.e.~an extra cost, with the big advantage of being a self-sufficient observable, involving only the transversity distribution (``squared''). All the others, while requiring a single polarized beam, imply, unavoidably, the appearance of extra unknown soft functions.

\subsection {Double transverse spin asymmetries}
Within the first type of DtSAs a major role is played by the Drell-Yan (DY) process with transversely polarized protons, $p^\uparrow p^\uparrow\to l^+l^-\,X$, as proposed in the seminal paper by Ralston and Soper~\cite{Ralston:1979ys}. The double spin asymmetry is given as
\be
A_{TT} \equiv \frac{d\sigma^{\uparrow\uparrow} -
d\sigma^{\uparrow\downarrow}} {d\sigma^{\uparrow\uparrow} +
d\sigma^{\uparrow\downarrow}} \sim \sum_q e_q^2 \left[ h_1^q(x_1) \,
h_1^{\bar q}(x_2) +  h_1^{\bar q}(x_1) \, h_1^q(x_2)\right]\,.
\ee
This measurement would be feasible at RHIC where large center of mass energies can be reached. On the other hand, the small $x$ region covered, together with the expected small values of $h_1^q$ for antiquarks, give an upper bound for $A_{TT}$ of 1-2\%~\cite{Martin:1999mg}.
Much larger asymmetries are expected by using transversely polarized antiprotons ($p^\uparrow \bar p^\uparrow$), as proposed by the PAX Collaboration~\cite{Barone:2005pu}, thanks to the lower energy and the fact that $A_{TT}$ is given in terms of the product of two quark transversity distributions. Unfortunately the polarization of antiprotons is still a formidable task and too low rates are expected. That is why, to overcome this difficulty, it has been proposed to look at the $J/\psi$ peak (gaining a factor two in statistics)~\cite{Anselmino:2004ki}. Other DtSAs, like those for inclusive photon, jet or pion production, are strongly suppressed due to the gluon dominance in their denominator.

For processes with a final polarized hadron, lambda production in semi-inclusive deep inelastic scattering (SIDIS), or in $pp$ collisions, could be extremely helpful. Its main advantage is the self-analyzing power of $\Lambda$ via its parity violating decay. The price is that the spin transfer, $D_{NN}$, is given in terms of $h_1^q$ coupled to the unknown transversely polarized fragmentation function $H_1^q$. Moreover, in SIDIS, the dominance of $u$ quarks in the proton, together with the expected relevant role of the fragmenting strange quark in the spin transfer to the $\Lambda^\uparrow$, implies low values for $D_{NN}$.
Complementary and necessary information on $H_1^q$ can be obtained, for instance, from the study of $e^+e^-\to\Lambda^\uparrow \bar \Lambda^\uparrow X$~\cite{Contogouris:1995xc}.

\subsection{Single spin asymmetries}

$\bullet$ {$\Delta_T q$ via TMDs}\\
Again we can consider the partner in the initial state, like in $p^\uparrow p\to l^+l^-\,X$ (DY), or in the final state, like in $l p^\uparrow\to l'h\,X$ (SIDIS). These are the processes where TMD factorization has been proved to hold~\cite{Ji:2004wu,Ji:2004xq} and where one is able, by looking at specific azimuthal dependences, to disentangle unambiguously the terms involving $\Delta_T q$. More precisely, in DY one can access the transversity coupled with $h_1^{\perp q}$, the Boer-Mulders function~\cite{Boer:1999mm}, giving the probability to find a transversely polarized quark inside an unpolarized proton. This could be then extracted from the study of the $\cos2\phi$ dependence in the unpolarized DY cross section (involving $h_1^{\perp q}$ squared).
In SIDIS, with a transversely (T) polarized target, by measuring the azimuthal dependence of the final hadron w.r.t.~the lepton scattering plane, 
a peculiar modulation involving $h_1^q$ emerges ($\phi_S$, $\phi_h$ being the azimuthal angles, respectively,  of the proton spin and the final hadron momentum)~\cite{Mulders:1995dh, Bacchetta:2006tn,Anselmino:2011ch}
\be
A_{UT}\sim \cdots + h_1^q\otimes H_1^{\perp q} \sin (\phi_h+\phi_S)\,.
\ee
The extra unknown, $H_1^{\perp q}$, is the Collins function~\cite{Collins:1992kk}, giving the probability for a transversely polarized quark to fragment into an unpolarized hadron. It can be extracted from the azimuthal asymmetries in the distribution of two almost back-to-back hadrons in $e^+e^-$ annihilation (involving $H_1^{\perp q}$ squared)~\cite{Boer:1997mf}.
The experimental evidence of these asymmetries~\cite{Airapetian:2004tw,Abe:2005zx} has indeed allowed for the first-ever extraction of $h_1^q$~\cite{Anselmino:2007fs}.

Another possible source of information on transversity, still within a TMD factorization scheme (even if not formally proven), is the study of the azimuthal distribution of a pion inside a jet in $p^\uparrow p$ collisions~\cite{Yuan:2007nd, D'Alesio:2010am}.

A word of caution on the analysis of SSAs within the TMD approach is mandatory. In general, beyond the tree-level approximation, TMD factorization involves an extra soft factor, that, depending on the scale, implies a dilution of the asymmetry at large $Q^2$~\cite{Boer:2001he, Boer:2008fr}. Some developments on TMD evolution have recently appeared~\cite{Aybat:2011ge,Anselmino:2012aa,GarciaEchevarria:2011rb}, but its quantitative effect (in particular for chiral-odd functions, like $H_1^{\perp q}$) is still under investigation and not yet taken into account in phenomenological extractions.\\

$\bullet$ {$\Delta_Tq$ via DiFFs}\\
In the process $lp^\uparrow\to l' (\pi\pi) \, X$, the quark fragmentation mechanism $q^\uparrow\to \pi\pi$ acts as a polarimeter and one can access $h_1^q$ in combination with the dihadron fragmentation function $H_1^\open$~\cite{Jaffe:1997hf, Radici:2001na}, describing the correlation between the transverse polarization of the fragmenting quark and the azimuthal orientation of the plane containing the momenta of the detected hadron pair.
Again this extra unknown must be extracted by analyzing, for instance, double azimuthal correlations in $e^+e^-\to (\pi\pi)_1 (\pi\pi)_2\, X$~\cite{Artru:1995zu}. The main advantage of this method is that is based on the standard collinear factorization and the evolution of DiFFs is known to be the same as that of $H_1^q$.

\begin{figure}[b!]
\centering
\begin{minipage}[c]{.4\textwidth}
\centering
\includegraphics[width=1\textwidth, angle=-90]{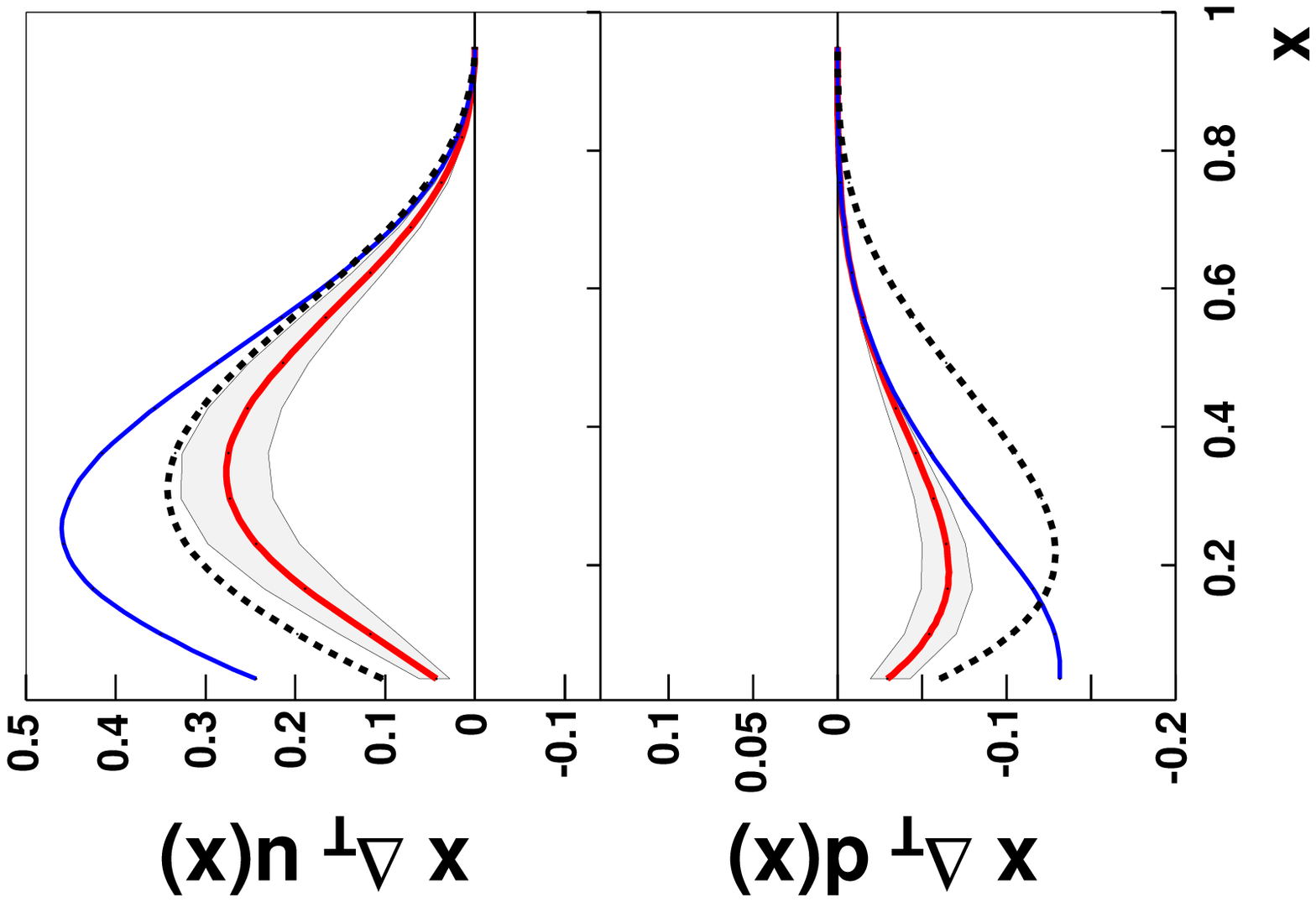}
\end{minipage}%
\hspace{2mm}%
\begin{minipage}[c]{.4\textwidth}
\centering
\includegraphics[width=1\textwidth]{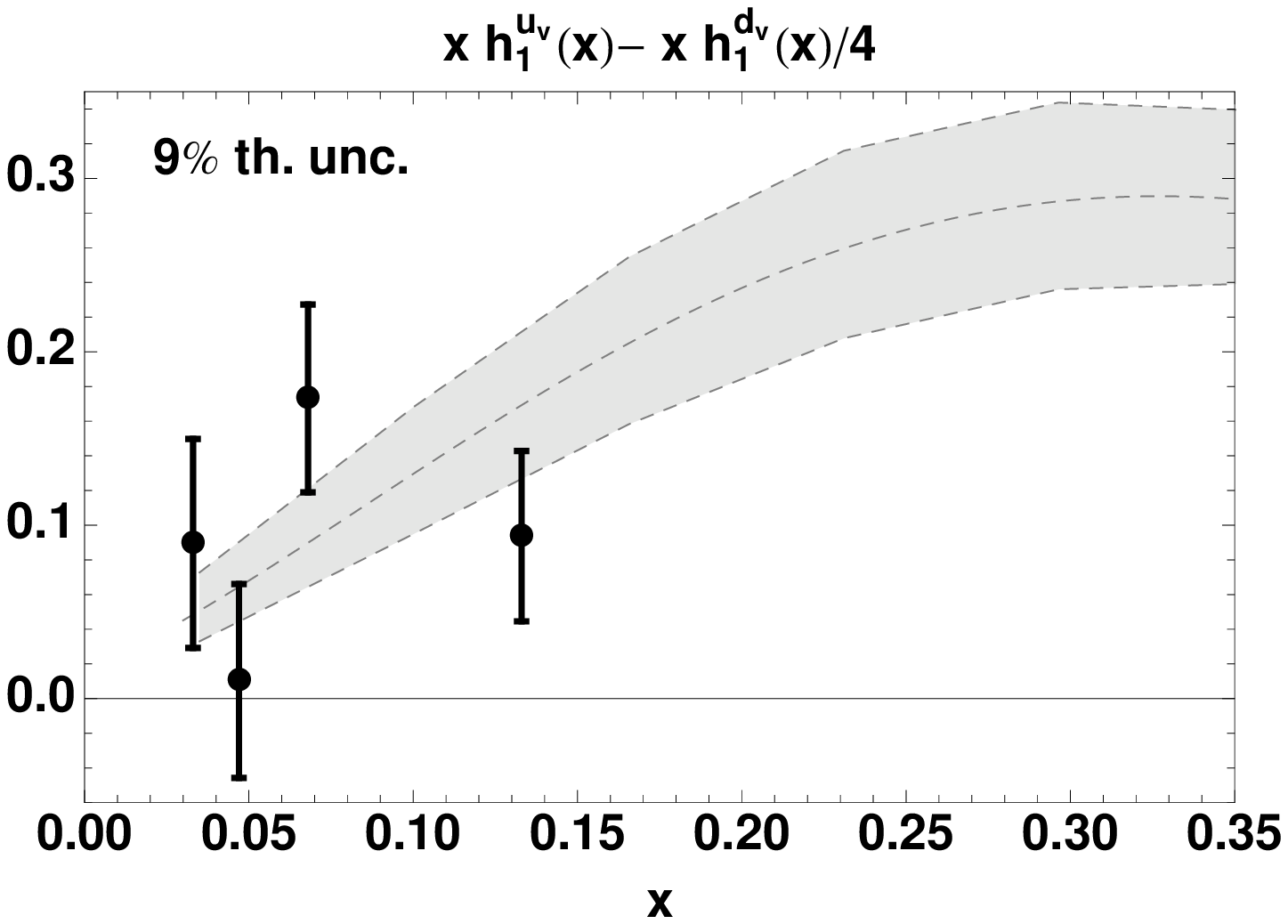}
\end{minipage}
\caption{Left panel: $\Delta_Tq$ (solid red line) from the fit~\cite{Anselmino:2008jk} on the Collins effect. Helicity distribution (dotted black line), Soffer bound (solid blue line) and uncertainty band are also shown. Right panel: Results from Ref.~\cite{Bacchetta:2011ip} together with the uncertainty band as deduced from the fit on the left panel.
}
\end{figure}

\section{Present status and perspectives}
Thanks to formidable efforts of various experimental groups, HERMES~\cite{Airapetian:2004tw, Airapetian:2010ds} and \cite{Airapetian:2008sk}, COMPASS~\cite{Alexakhin:2005iw, Alekseev:2010rw}, and Belle~\cite{Abe:2005zx,Seidl:2008xc} and \cite{Vossen:2011fk}, large data sets have become available, allowing, at last, for two independent phenomenological extractions of $\Delta_T q$.

Concerning the analysis of the Collins effect~\cite{Anselmino:2007fs, Anselmino:2008jk},
the main aspects are: $i)$ a power-like ($x^a(1-x)^b$) parametrization of $h_1^q$ for up and down quarks and, similarly, of the favored ($u\to\pi^+$) and unfavored ($d\to\pi^+$) Collins functions, with a factorized gaussian $k_\perp$ dependence; $ii)$ use of the universality property of $H_1^{\perp q}$~\cite{Metz:2002iz, Collins:2004nx}; $iii)$ simultaneous extraction of $h_1^q$ and $H_1^\perp$ from a global fit of SIDIS and $e^+e^-$ data; $iv)$ $Q^2$-evolution properly taken into account for $h_1^q$, while replaced for $H_1^{\perp q}$ with that of the unpolarized fragmentation function. The results~\cite{Anselmino:2008jk} are shown in Fig.~1 (left panel), where one can see how $\Delta_T q$ (solid red line), at $Q^2=2.4$ GeV$^2$, is sizeable, different from $\Delta q$ (dotted black line) and much smaller than the Soffer bound~\cite{Soffer:1994ww} (solid blue line).

For the latest complementary study via DiFFs~\cite{Bacchetta:2011ip} the main features are: $i)$ extraction of $H_1^\open$ for $(\pi^+\pi^-)$ from $e^+e^-$ data; $ii)$ $H_1^{\open\, u} = - H_1^{\open\, d}$; $iii)$ unpolarized $D^{q\to\pi\pi}$ from PITHYA, due to the absence of data; $iv)$ proper $Q^2$-evolution of $H_1^\open$ from 110 to 2.4 GeV$^2$; $v)$~extraction from SIDIS data of the combination $(xh_1^u-xh_1^d/4)$.
The results of this fit, with their statistical errors, are presented in Fig.~1 (right panel), where the uncertainty band of the fit via TMDs is also shown. The compatibility of the two extractions, taking into account the various assumptions behind them, is quite encouraging.

Summarizing, transversity has become definitely a hot topic in spin physics: theoretically well known, and, recently, accessed also phenomenologically. DtSAs, more difficult experimentally, are much cleaner from the theoretical point of view, with $A_{TT}$ in DY processes being the golden channel.
Concerning SSAs, improvements in the proper $Q^2$-evolution of TMDs will be of great help in view of global fits, as well as more data (in particular at JLab in the still unexplored large $x$ region) are eagerly awaited.


\begin{theacknowledgments}
The author thanks the organizers of this very nice conference for their kind invitation.
\end{theacknowledgments}



\bibliographystyle{aipproc}   


\begin{thebibliography}{39}
\expandafter\ifx\csname natexlab\endcsname\relax\def\natexlab#1{#1}\fi
\providecommand{\enquote}[1]{``#1''}
\expandafter\ifx\csname url\endcsname\relax
  \def\url#1{\texttt{#1}}\fi
\expandafter\ifx\csname urlprefix\endcsname\relax\def\urlprefix{URL }\fi
\providecommand{\eprint}[2][]{\url{#2}}

\bibitem[Barone et~al.(2002)]{Barone:2001sp}
V.~Barone, A.~Drago, and P.~G. Ratcliffe, \emph{Phys.Rep.} \textbf{359},
  1--168 (2002).

\bibitem[Soffer(1995)]{Soffer:1994ww}
J.~Soffer, \emph{Phys.Rev.Lett.} \textbf{74}, 1292--1294 (1995).

\bibitem[Gockeler et~al.(2007)]{Gockeler:2006zu}
M.~Gockeler, et~al., \emph{Phys.Rev.Lett.} \textbf{98}, 222001 (2007).

\bibitem[Ralston and Soper(1979)]{Ralston:1979ys}
J.~P. Ralston, and D.~E. Soper, \emph{Nucl.Phys.} \textbf{B152}, 109 (1979).

\bibitem[Martin et~al.(1999)]{Martin:1999mg}
O.~Martin, A.~Schafer, M.~Stratmann, and W.~Vogelsang, \emph{Phys.Rev.}
  \textbf{D60}, 117502 (1999).

\bibitem[Barone et~al.(2005)]{Barone:2005pu}
V.~Barone, et~al.  (2005), \eprint{hep-ex/0505054}.

\bibitem[Anselmino et~al.(2004)]{Anselmino:2004ki}
M.~Anselmino, V.~Barone, A.~Drago, and N.~N. Nikolaev, \emph{Phys.Lett.}
  \textbf{B594}, 97--104 (2004).

\bibitem[Contogouris et~al.(1995)]{Contogouris:1995xc}
A.~Contogouris, O.~Korakianitis, Z.~Merebashvili, and F.~Lebessis,
  \emph{Phys.Lett.} \textbf{B344}, 370--376 (1995).

\bibitem[Ji et~al.(2005)]{Ji:2004wu}
X.-d. Ji, J.-p. Ma, and F.~Yuan, \emph{Phys.Rev.} \textbf{D71}, 034005
  (2005).

\bibitem[Ji et~al.(2004)]{Ji:2004xq}
X.-d. Ji, J.-P. Ma, and F.~Yuan, \emph{Phys.Lett.} \textbf{B597}, 299--308
  (2004).

\bibitem[Boer(1999)]{Boer:1999mm}
D.~Boer, \emph{Phys.Rev.} \textbf{D60}, 014012 (1999).

\bibitem[Mulders and Tangerman(1996)]{Mulders:1995dh}
P.~Mulders, and R.~Tangerman, \emph{Nucl.Phys.} \textbf{B461}, 197--237
  (1996).

\bibitem[Bacchetta et~al.(2007)]{Bacchetta:2006tn}
A.~Bacchetta, et~al., \emph{JHEP} \textbf{02}, 093 (2007).

\bibitem[Anselmino et~al.(2011)]{Anselmino:2011ch}
M.~Anselmino, et~al., \emph{Phys.Rev.} \textbf{D83}, 114019 (2011).

\bibitem[Collins(1993)]{Collins:1992kk}
J.~C. Collins, \emph{Nucl.Phys.} \textbf{B396}, 161--182 (1993).


\bibitem[Boer et~al.(1997)]{Boer:1997mf}
D.~Boer, R.~Jakob, and P.~J. Mulders, \emph{Nucl.Phys.} \textbf{B504},
  345--380 (1997).

\bibitem[Airapetian et~al.(2005)]{Airapetian:2004tw}
A.~Airapetian, et~al., \emph{Phys.Rev.Lett.} \textbf{94}, 012002 (2005).


\bibitem[Abe et~al.(2006)]{Abe:2005zx}
K.~Abe, et~al., \emph{Phys.Rev.Lett.} \textbf{96}, 232002 (2006).

\bibitem[Anselmino et~al.(2007)]{Anselmino:2007fs}
M.~Anselmino, et~al., \emph{Phys.Rev.} \textbf{D75}, 054032 (2007).

\bibitem[Yuan(2008)]{Yuan:2007nd}
F.~Yuan, \emph{Phys.Rev.Lett.} \textbf{100}, 032003 (2008).

\bibitem[D'Alesio et~al.(2011)]{D'Alesio:2010am}
U.~D'Alesio, F.~Murgia, and C.~Pisano, \emph{Phys.Rev.} \textbf{D83}, 034021
  (2011).

\bibitem[Boer(2001)]{Boer:2001he}
D.~Boer, \emph{Nucl.Phys.} \textbf{B603}, 195--217 (2001).

\bibitem[Boer(2009)]{Boer:2008fr}
D.~Boer, \emph{Nucl.Phys.} \textbf{B806}, 23--67 (2009).


\bibitem[Aybat et~al.(2012)]{Aybat:2011ge}
S.~M. Aybat, J.~C. Collins, J.-W. Qiu, and T.~C. Rogers, \emph{Phys.Rev.}
  \textbf{D85}, 034043 (2012).

\bibitem[Anselmino et~al.(2012)]{Anselmino:2012aa}
M.~Anselmino, M.~Boglione and S.~Melis, \emph{Phys.Rev.}
  \textbf{D86}, 014028 (2012).

\bibitem[GarciaEchevarria et~al.(2011)]{GarciaEchevarria:2011rb}
M.~Garcia-Echevarria, A.~Idilbi, and I.~Scimemi,  \emph{JHEP}
  \textbf{1207}, 002 (2012).

\bibitem[Jaffe et~al.(1998)]{Jaffe:1997hf}
R.~L. Jaffe, X.-m. Jin, and J.~Tang, \emph{Phys.Rev.Lett.} \textbf{80},
  1166--1169 (1998).

\bibitem[Radici et~al.(2002)]{Radici:2001na}
M.~Radici, R.~Jakob, and A.~Bianconi, \emph{Phys.Rev.} \textbf{D65}, 074031
  (2002).

\bibitem[Artru and Collins(1996)]{Artru:1995zu}
X.~Artru, and J.~C. Collins, \emph{Z.Phys.} \textbf{C69}, 277--286 (1996).

\bibitem[Airapetian et~al.(2010)]{Airapetian:2010ds}
A.~Airapetian, et~al., \emph{Phys.Lett.} \textbf{B693}, 11--16 (2010).

\bibitem[Airapetian et~al.(2008)]{Airapetian:2008sk}
A.~Airapetian, et~al., \emph{JHEP} \textbf{0806}, 017 (2008).

\bibitem[Alexakhin et~al.(2005)]{Alexakhin:2005iw}
V.~Y. Alexakhin, et~al., \emph{Phys.Rev.Lett.} \textbf{94}, 202002 (2005).

\bibitem[Alekseev et~al.(2010)]{Alekseev:2010rw}
M.~Alekseev, et~al., \emph{Phys.Lett.} \textbf{B692}, 240--246 (2010).

\bibitem[Seidl et~al.(2008)]{Seidl:2008xc}
R.~Seidl, et~al., \emph{Phys.Rev.} \textbf{D78}, 032011 (2008).

\bibitem[Vossen et~al.(2011)]{Vossen:2011fk}
A.~Vossen, et~al., \emph{Phys.Rev.Lett.} \textbf{107}, 072004 (2011).

\bibitem[Anselmino et~al.(2009)]{Anselmino:2008jk}
M.~Anselmino, et~al., \emph{Nucl.Phys.Proc.Suppl.} \textbf{191}, 98--107
  (2009).

\bibitem[Metz(2002)]{Metz:2002iz}
A.~Metz, \emph{Phys.Lett.} \textbf{B549}, 139--145 (2002).

\bibitem[Collins and Metz(2004)]{Collins:2004nx}
J.~C. Collins, and A.~Metz, \emph{Phys.Rev.Lett.} \textbf{93}, 252001 (2004).

\bibitem[Bacchetta et~al.(2011)]{Bacchetta:2011ip}
A.~Bacchetta, A.~Courtoy, and M.~Radici, \emph{Phys.Rev.Lett.} \textbf{107},
  012001 (2011).

\end{thebibliography}

\hyphenation{Post-Script Sprin-ger}

\end{document}